\documentclass[10pt,conference,a4paper]{IEEEtran}

\usepackage{float}

\usepackage[pdftex, pagebackref=true, pdfpagelabels=true,
hypertexnames=true, plainpages=false, naturalnames=false,
bookmarks=true, bookmarksnumbered=true,
pdfpagemode=UseOutlines,]{hyperref}

\hypersetup{colorlinks, citecolor=[rgb]{0,0.7,0}, filecolor=[rgb]{0.6,0,0}, linkcolor=[rgb]{0.6,0,0}, urlcolor=[rgb]{0.6,0,0}}
 
\usepackage{graphicx} \graphicspath{{figures/}}

\usepackage[first=0,last=9]{lcg}

\usepackage{amsmath}
\usepackage{booktabs}       % professional-quality tables
\usepackage{amsfonts}       % blackboard math symbols
\usepackage{nicefrac}       % compact symbols for 1/2, etc.
\usepackage{multirow}
\usepackage{microtype}
\usepackage{authblk}
\usepackage{balance}

\usepackage{algorithm, algorithmic}
\usepackage[algo2e]{algorithm2e}

\usepackage{color, colortbl, xcolor}
\definecolor{LightCyan}{rgb}{0.88,1,1}
\newcolumntype{?}{!{\vrule width 1pt}}
\newcolumntype{a}{>{\columncolor{lightgray}}c}

\begin{document}

\title{Progressive Adversarial Semantic Segmentation}

\author{Abdullah-Al-Zubaer Imran}
\author{Demetri Terzopoulos}
\affil{Computer Science Department\\University of California, Los Angeles, California, USA}
\maketitle

\begin{abstract}
Medical image computing has advanced rapidly with the advent of deep learning techniques such as convolutional neural networks. Deep convolutional neural networks can perform exceedingly well given full supervision. However, the success of such fully-supervised models for various image analysis tasks (e.g., anatomy or lesion segmentation from medical images) is limited to the availability of massive amounts of labeled data. Given small sample sizes, such models are prohibitively data biased with large domain shift. To tackle this problem, we propose a novel end-to-end medical image segmentation model, namely Progressive Adversarial Semantic Segmentation (PASS), which can make improved segmentation predictions without requiring any domain-specific data during training time. Our extensive experimentation with 8 public diabetic retinopathy and chest X-ray datasets, confirms the effectiveness of PASS for accurate vascular and pulmonary segmentation, both for in-domain and cross-domain evaluations.          
\end{abstract}

\begin{IEEEkeywords}
segmentation, adversarial learning, domain-shift, diabetic retinopathy, chest X-ray, lung, retinal vessels
\end{IEEEkeywords}

\section{Introduction} 

%Why segmentation and the pitfalls of iid assumption
% Deep learning has become extremely successful in exploiting the data-driven models for various image analysis tasks across computer vision and medical imaging. 
% Medical image segmentation is one of the most important tasks that allows analyze any object of interest such as, anatomical structures, lesions, etc. 
Deep learning-based image segmentation methods perform well amid the availability of large pools of labeled data.
%\citep{imran2018automatic}. 
An important assumption underlying the success of deep convolutional neural networks is that the training data are sampled iid (independent, identical distribution). However, in real world scenarios, that assumption may not hold. By breaking the iid assumption, deep learning models can achieve robustness across different data domains. 

\begin{figure}[t]
    \centering
    \resizebox{\linewidth}{!}{
    \begin{tabular}{l | l}
    \includegraphics[width=\linewidth, trim={2.5cm 1.25cm 2cm 1cm}, clip]{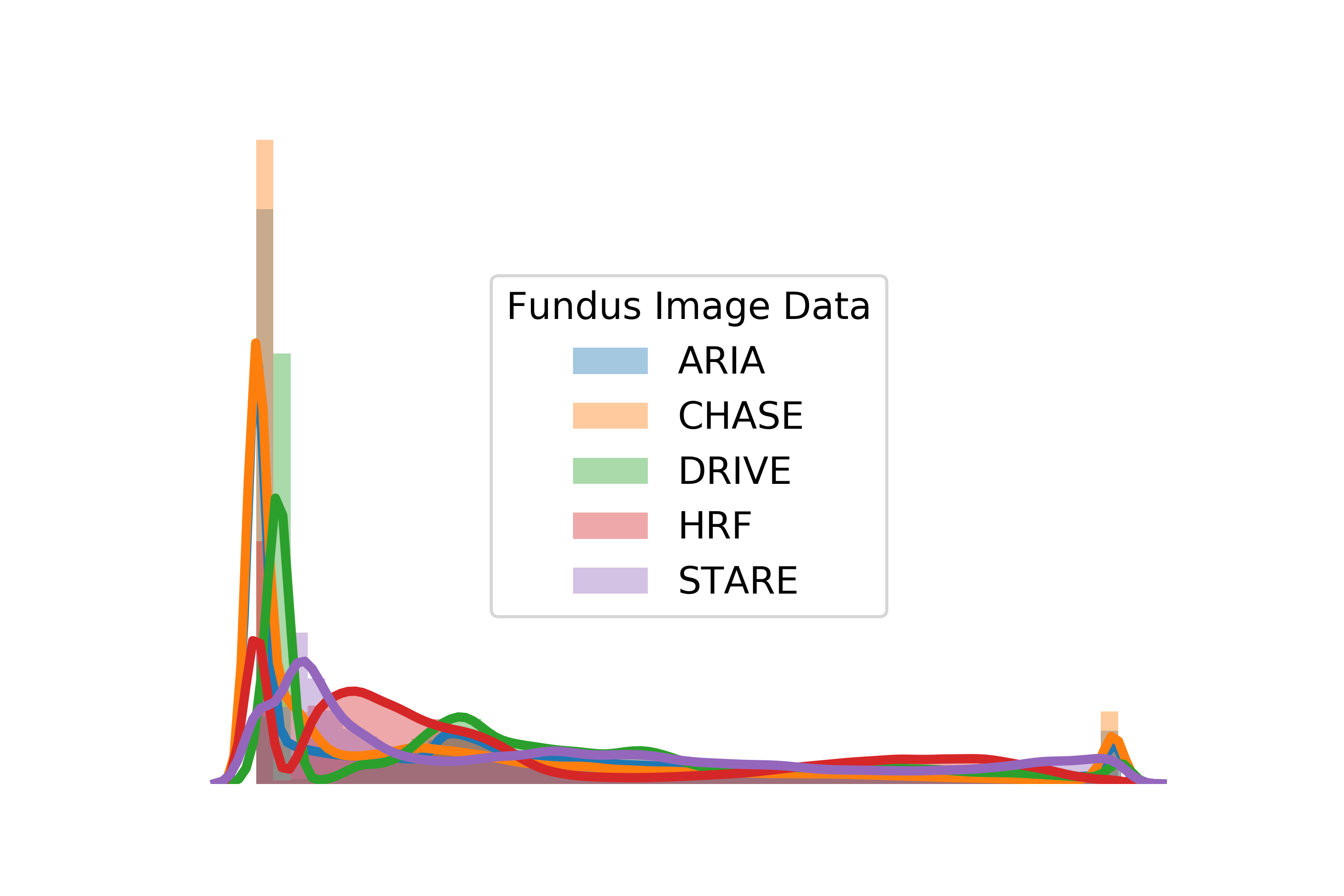}
    &
    \includegraphics[width=\linewidth, trim={2.25cm 1.25cm 2cm 1cm}, clip]{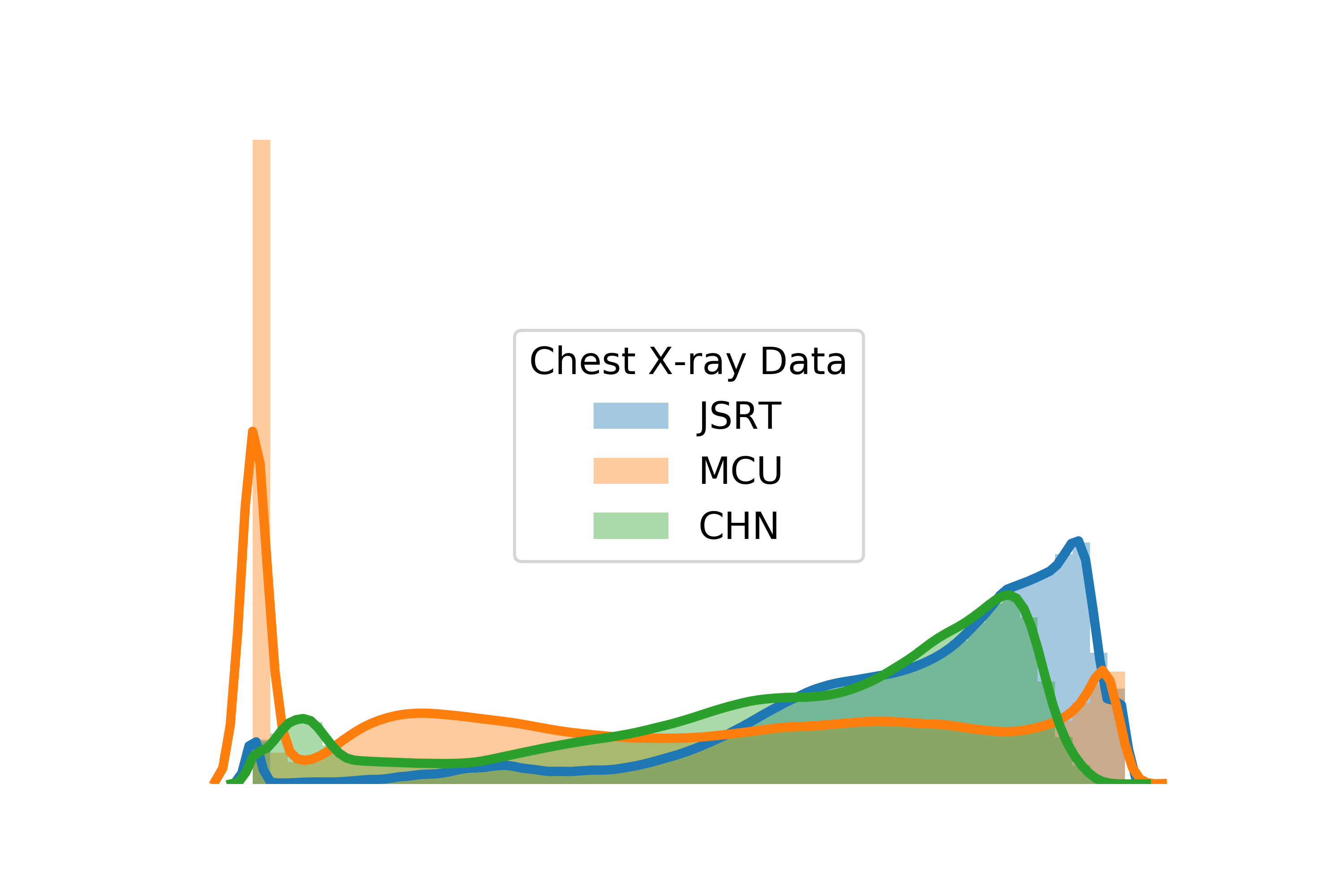}
    \end{tabular}
    }
    \caption{Disparity in distributions of the datasets used in our experiments.}
    \label{fig:data_dist}
\end{figure}

Unlike the availability of natural image datasets, where the success of deep learning is mainly due to large training set \cite{krizhevsky2012imagenet}, due to privacy issues and the scarcity of manual annotations, medical imaging datasets available for training are usually modest in size.  Moreover, the problem is exacerbated due to the under-representation of some rare medical conditions, varying imaging configurations, modalities, among other factors (Fig.~\ref{fig:data_dist}). This leads to the domain shift problem---when a model trained on data from one source does not generalize well or fails to predict proper segmentation for data from a different source. In that case, very non-intuitive and error-prone segmentations can be generated, even by the most sophisticated deep learning models \cite{imran2018automatic, zhou2019unet++}. Thus training CNNs on medical imaging data for segmentation tasks lacks robustness when the iid assumption breaks, and they make terrible mistakes for ood (out of distribution) predictions.

Domain shift is usually tackled through unsupervised domain adaptation, where cross-domain data are required at training time \cite{dou2019unsupervised, chen2020unsupervised, dou2018unsupervisedcross}. Domain generalization is another approach that tries to learn a common representation for the data from both source and target domains. However, the additional unlabeled data required from the target domain are not readily available. Moreover, such approaches are not clinically viable due to their scaling issues across many target domains. A cost-effective solution is interactive segmentation, guided by experts in the loop. However, because of the manual intervention in the segmentation process, these approaches are not methodically feasible \cite{sakinis2019interactive}. An end-to-end and elegant solution would not rely on manual intervention nor require training data from target domains. 

\begin{figure}[b]
    \centering
    \resizebox{\linewidth}{!}{
    \begin{tabular}{c c c c}
    \includegraphics[width=\linewidth, trim={0cm 6.25cm 21cm 6.25cm}, clip]{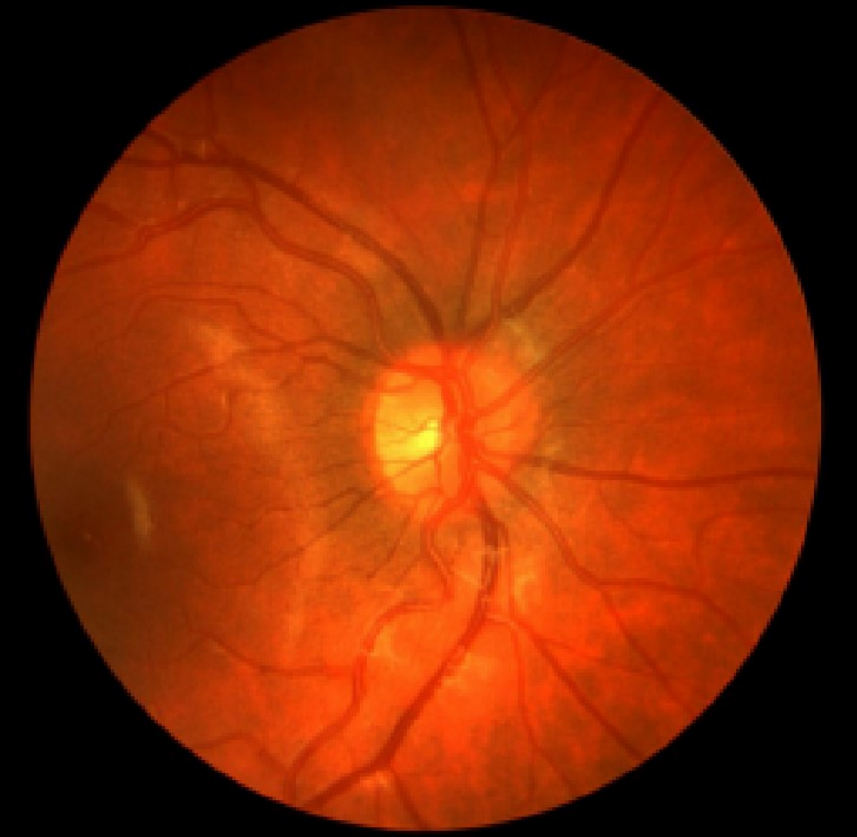}
     &
    \includegraphics[width=\linewidth, trim={10.5cm 6.25cm 10.5cm 6.25cm}, clip]{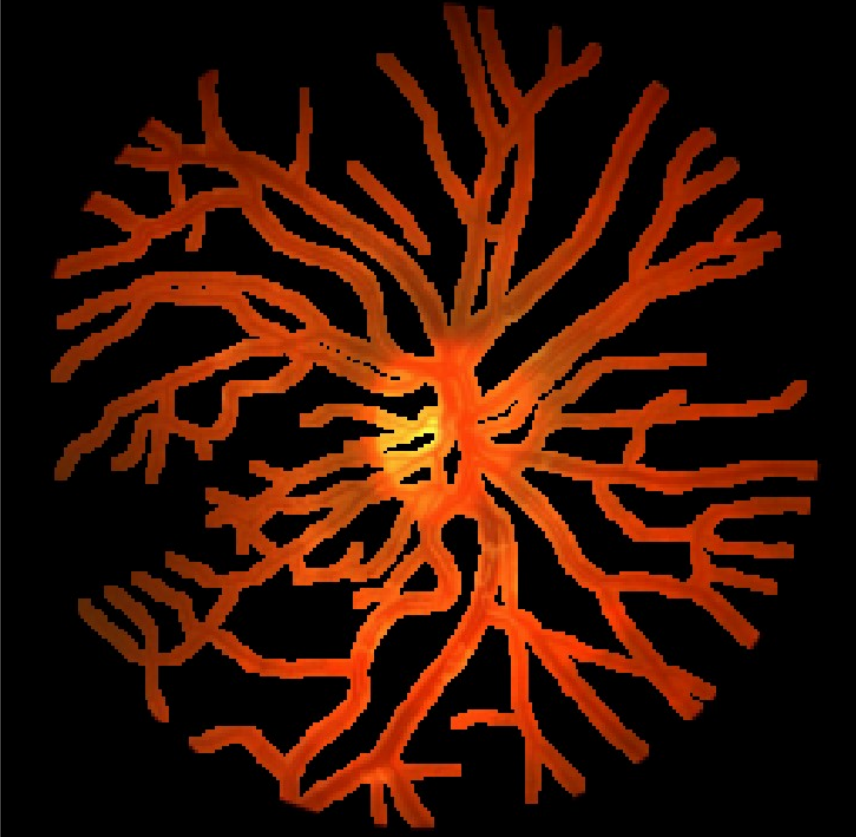}
    &
    \includegraphics[width=\linewidth, trim={21cm 6.25cm 0cm 6.25cm}, clip]{chase_003_aria.pdf}
    &
    \includegraphics[width=\linewidth, trim={21cm 6.25cm 0cm 6.25cm}, clip]{chase_003_chase.pdf}
    \\
    \smallskip
    {\Huge Fundus Image } & {\Huge Ground Truth} & {\Huge ARIA} & {\Huge CHASE}
    \end{tabular}
    }\\[6pt]
    \resizebox{\linewidth}{!}{
    \begin{tabular}{c c c c}
    \includegraphics[width=\linewidth, trim={4.75cm 0.25cm 4.75cm 1cm}, clip]{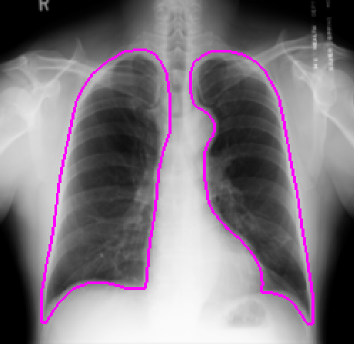}
    &
    \includegraphics[width=\linewidth, trim={9.25cm 0.25cm 0.25cm 1cm}, clip]{nlm_bound_031_nlm.pdf}
        &
    \includegraphics[width=\linewidth, trim={9.25cm 0.25cm 0.25cm 1cm}, clip]{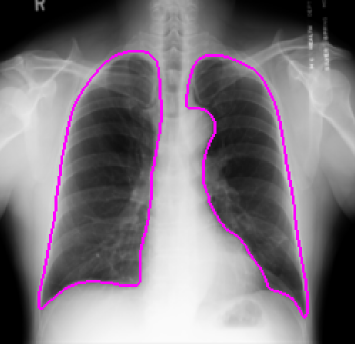}
    &
    \includegraphics[width=\linewidth, trim={9.25cm 0.25cm 0.25cm 1cm}, clip]{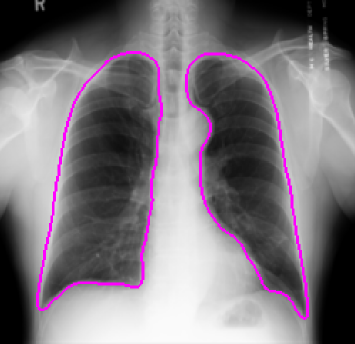}
    \\
    \smallskip
    {\Huge Ground Truth} & {\Huge MCU (98.99)} & {\Huge JSRT (97.27)} & {\Huge CHN (98.60)}
    \end{tabular}
    }
    \caption{Consistency of the in-domain and cross-domain segmentation predictions by our PASS model: (Top) Visualization of retinal vessel segmentation from a fundus image when trained on the ARIA and CHASE datasets. (Bottom) Segmentation of a chest X-ray from the MCU dataset when the model is trained on the MCU, JSRT, and CHN datasets.}
    \label{fig:teaser}
\end{figure}

\begin{figure*}
    \centering
    \includegraphics[width=0.78\linewidth, trim={1cm 0cm 0.25cm 0cm}, clip]{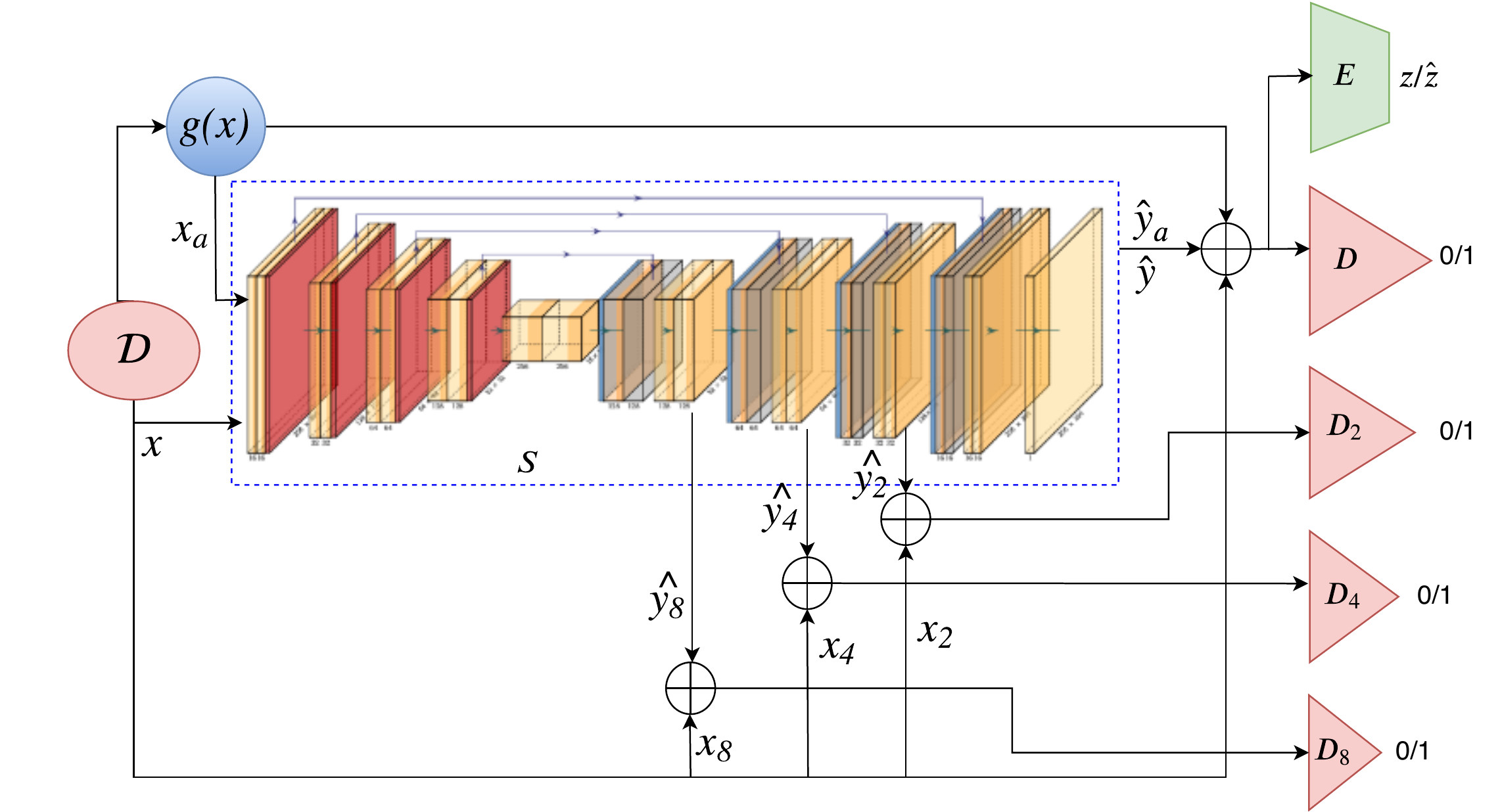}
    \caption{Schematic of the PASS model: The segmentation mask generator $S$ takes either input $x$ or the transformed input $x_a$ via a transformation function $g(x)$. The generated side outputs are passed to the corresponding discriminators $D_2$, $D_4$, and $D_8$ and the final outputs $y_a/\hat{y}$ are passed to the discriminator $D$. The shape encoder $E$ also takes $y/\hat{y}$ as input to get the latent vector $z/\hat{z}$.}
    \label{fig:model}
\end{figure*}

{\bf Contributions:} We introduce Progressive Adversarial Semantic Segmentation, or PASS, a novel segmentation model that improves segmentation predictions by tackling domain shift and small sample size issues. Our specific contributions are: {\bf1)} A novel semantic segmentation model with progressive side-supervision and adversarial learning for improved, shape-aware, and anatomically-consistent segmentation of medical images. {\bf2)} Reduction of domain shift for state-of-the-art models in small training data, with improved segmentation prediction in both in-domain and cross-domain evaluations. {\bf3)} Extensive experimentation and statistical analyses with retinal fundus and chest X-ray image datasets, demonstrating the effectiveness of the PASS model in vascular and pulmonary segmentation (Fig.~\ref{fig:teaser}).

\section{Related Work}
% There are some methods available in literature on dealing with limited labeled dataset and domain shift problems. 

The following five topics are relevant to the present work: semi-supervised learning, adversarial learning, domain adaptation, prior-/post-processing segmentation, and error correction. 

Semi-supervised models are evaluated by retaining only a portion of the labels from a dataset while the remainder are treated as unlabeled data \cite{zhai2019s}. Semi-supervised learning can be performed in at least two different ways---self-supervised and adversarial learning. Tajbakhsh {\it et al.}~\cite{tajbakhsh2019surrogate} showed the effectiveness of training models from pre-trained surrogate tasks in lung lobe segmentation with limited labeled data. 
% \citep{imran2020self} proposed a semi-supervised learning scheme based on self-supervision branching fully supervised and self-supervised for joint medical image classification and segmentation. 
Adversarial learning can effectively be adapted to semi-supervised learning for the classification of medical images \cite{salehinejad2018generalization}. Adversarial learning has also been utilized in segmentation (semantic-aware generative adversarial nets \cite{chen2018semantic}, structure correcting adversarial nets \cite{dai2018scan}, etc.). We proposed a semi-supervised multitasking model with adversarial training \cite{imran2019semi}. 
Domain adaption can also be performed to learn a generic representation where the model is fully-supervised for source data and unsupervised for the target data \cite{yang2019domain}. These methods are typically reliant on the availability of unlabeled \cite{chen2019synergistic, huo2018synseg} or even labeled data \cite{zhang2018translating, dou2018pnp} for the target domain. Therefore, with the unavailability of labeled/unlabeled data from the target domains or disparate data distributions, such models become less useful. 
Post-processing approaches inconsistently improve segmentation results but require extensive parameter tuning \cite{kamnitsas2017efficient} and error propagations before the post-processing \cite{larrazabal2019anatomical}. ErrorNet is another method proposed recently that can learn error correction in a systematic manner via learning a prior distribution of the segmentation masks \cite{tajbakhsh2019errornet}. 
However, all of these methods are based on explicit error propagation whether it is handcrafted or model-derived. This adds vulnerability to the models for segmentation prediction and might increase shifts across different application tasks.

Departing from all the previous methods, PASS is end-to-end and fully automatic, devoid of any pre- or post-processing and explicit error-designs (handcrafted/systematic),  and, most importantly, better generalized for filling domain gaps.

\section{PASS}

% PASS is the progressive adversarial semantic segmentation which exploits side-supervision against multi-scale discriminators in an adversarial learning framework. With a shape encoder and local logit-wise distribution matching, the model is enabled for accurate segmentation prediction in out-of-sample-distributions. \citet{imran2018automatic}.

To formulate the problem, we assume an unknown data distribution $p(X,Y)$ over images and segmentation labels. The model has access to the labeled training set $\mathcal{D}$. and unlabeled set $\mathcal{D_A}$ through on-the-fly transformation from $p(X)$ after marginalizing out $Y$. We set the learning objectives for the segmentation task as:
\begin{equation}
    \min_{\psi, \phi, \theta}\ \mathcal{L_L}(\mathcal{D}, (\psi, \phi, \theta)) + \lambda\mathcal{L_A}(\mathcal{D_A}, (\psi, \phi, \theta)), 
    \end{equation}
where the supervised objective $\mathcal{L_L}$ is defined on the labeled data and the unsupervised objective $\mathcal{L_A}$ is defined on the unlabeled data, $\lambda$ is a non-negative weight parameter, and $\psi$, $\phi$, and $\theta$ denote the parameters of the segmentor $S$, discriminator $D$, and encoder $E$ networks, respectively.

%%%%%%%%%%%%%%%%%%%%%%%Algorithm%%%%%%%%%%%%%%%%%%%%%%%%%%%%%%
\begin{algorithm}[t]
\caption{PASS Training.}
\label{alg:train}
\begin{algorithmic}
\small
\REQUIRE 
\STATE Training set of labeled data $x, y \in \mathcal{D}$ 
\STATE Transformation function $g(x)$ to generate $x_a$ from $x$ 
\STATE Network architecture $S_\phi, D_\psi. E_\theta \in \mathcal{F_{(\phi, \psi, \theta)}}$ with learnable parameters $\mathcal{\phi, \psi, \theta}$
\vspace{4pt}
\FOR{each {\bf epoch} over $\mathcal{D}$}
\STATE Generate minibatches of unlabeled inputs $\mathcal{M_A}$ using $g(x)$:\\
$x_a = g(x)$
\vspace{4pt}
\FOR{each {\bf step}}
\STATE Sample minibatch $\mathcal{M}$: $x_{(i)};{x_{(1)},\dots,x_{(m)}} \sim p_{\mathcal{D}(x)}$

\vspace{4pt}
Compute model outputs for the labeled inputs:\\ $\hat{y}\leftarrow \mathcal{F_(\phi,\psi,\theta}(\mathcal{M})$\\
Compute model outputs for the unlabeled inputs:\\ $\hat{y}_a\leftarrow \mathcal{F_\phi,\psi,\theta}(\mathcal{M_A})$

\vspace{4pt}
\STATE Update the discriminators $D_i (i=1,\dots,d)$ along their gradients:
\begin{align*}
    \nabla_{\psi_{D_i}} &\frac{1}{|\mathcal{M}|}\sum_{i\in\mathcal{M}}\left[L_{D_{i}{\left(x_{(i)}, y_{(i)}, \hat{y}_{(i)}\right)}}\right] + \\
    &\alpha \frac{1}{|\mathcal{M_A}|}\sum_{i\in\mathcal{M_A}}\left[L_{D_{i}{\left(x_{a_{(i)}},  \hat{y}_{a_{(i)}}\right)}}\right].
\end{align*}

\STATE Update the shape encoder $E$ along its gradient:
\begin{align*} \nabla_{\theta_{E}} &\frac{1}{|\mathcal{M}|}\sum_{i\in\mathcal{M}}\left[L_{E_{\left(x_{(i)}, y_{(i)}, \hat{y}_{(i)}\right)}}\right].
\end{align*}

\STATE Update the segmentation mask generator $S$ along its gradient:
\begin{align*} \nabla_{\phi_{S}} &\frac{1}{|\mathcal{M}|}\sum_{i\in\mathcal{M}}\left[L_{S_{\left(y_{(i)}, \hat{y}_{(i)}\right)}}\right] + \\
    &\alpha \frac{1}{|\mathcal{M_A}|}\sum_{i\in\mathcal{M_A}}\left[L_{S_{\left(y_{(i)}, \hat{y}_{a_{(i)}}\right)}}\right].
\end{align*}
\ENDFOR
\ENDFOR
\end{algorithmic}
\end{algorithm}
%%%%%%%%%%%%%%%%%%%%%%%%%%%%%%%%%%%%%%%%%%%%%%%%%%%%%%%%%%%%%%%%%%%

\subsection{Architecture Details}

Fig.~\ref{fig:model} illustrates the PASS model. PASS is based on the backbone of a progressive U-Net with some careful adjustments in the U-Net with side-adversary and side-supervision capabilities. As in a U-Net \cite{ronneberger2015u}, PASS has a segmentor $(S)$ with skip connections in an encoder-decoder architecture. 
In each encoder layer, two $3\times3$ convolutions are followed by instance normalization, leaky-ReLU activation, and a $2\times2$ max-pooling. We generate side-outputs in every stage of the decoder. The side-outputs are collected at the resolutions of $x/8$, $x/4$, and $x/2$ before the final output at the resolutions of $x$. According to the shape of the side-outputs, discriminators are employed and layers are added progressively. Progressively adding one side-output to the next improves the segmentation performance compared to collecting the output from the final decoder stage \cite{imran2018automatic}. The progressive side-outputs also ensure that the network does not lose track of objects of interest. Moreover, progressively growing the discriminators enables the model to receive feedback at different resolutions via the side-outputs. While the discriminators are employed at different side-outputs, the segmentor tries to generate side-outputs closer to the ground truths progressively for an improved and accurate final segmentation. A shape encoder $E$ is used in PASS to match the latent representation of the stacked input and output of $S$ with the stacked input and reference so that the model becomes shape-aware while mapping an input to the segmentation mask. Moreover, during training, a transformation function is used to obtain $x_a$ from input $x$, and PASS makes a segmentation prediction on $x_a$. Note that $x_a$ is used without any corresponding label information. Through this, PASS can be trained to make predictions on more diverse data crucial for evaluation in other domains.

\subsection{Loss Functions}

The three networks $S$, $D$, and $E$ in the PASS model, are trained on separate objectives, according to Algorithm~\ref{alg:train}:

\paragraph{Segmentor Loss} The objective of segmentor $S$ is based on the segmentation maps generated in different resolutions. We chose Dice loss to penalize the model for the segmentation map predictions. Therefore, the objective of $S$ includes segmentation loss as a weighted sum of all the side-output and side-adversarial losses, where $S$ wants the discriminators $D_i$ to maximize the likelihood for the predicted segmentations. For the segmentation predictions, we employ Dice losses and the final loss is calculated as $\mathcal{L}_{S_{seg}} = \sum_i^4w_i\mathcal{L}_{(y_i, \hat{y}_i)}$. 
\iffalse
\begin{equation}
\label{eqn:seg_loss} 
    \begin{aligned}
L_{seg} &= 1 - \frac{
  \sum_i^{m^2} y_{pl}^{(i)}\hat{y}_{pl}^{(i)} + \epsilon}{
  \sum_i^{m^2}y_{pl}^{(i)}\hat{y}_{pl}^{(i)} + \sum_i^{m^2} y_{p\bar{l}}^{(i)}\hat{y}_{pl}^{(i)} + \epsilon
  },
    \end{aligned}
\end{equation}
where, $\hat{y}_{pl}(i)$ is the predicted probability that the $i^{th}$-pixel is marked as foreground label and $\hat{y}_{p\bar{l}}(i)$ is the probability that the $i^{th}$-pixel is marked as background label. Similarly, $y_{pl}(i)$ and $y_{p\bar{l}}(i)$ denote pixel-wise mapping of the labels in the ground truth masks. 
\fi
A second segmentation loss term is used for logit-wise distribution comparison. Since $x_a$ is not paired with any reference segmentations, it is not possible to directly compare segmentation loss. Rather, we employ Kullback-Leibler (KL) divergence to penalize $S$ for not maintaining the distribution of the predicted segmentation of the labeled data. The KL loss is calculated as
\begin{equation}
    \label{eqn:seg_u_kl}
    \mathcal{L}_{S_{\textrm{KL}}} = \sum_i^{m^2}\left| (\hat{y}_{pk}{(i)} - \hat{y}_{a_{pk}}{(i)})\log({y_{pk}{(i)}}/{\hat{y}_{a_{pk}}{(i)}})\right|.
\end{equation}
Then, the segmentor's adversarial loss is calculated from the stacked input image and predicted segmentation when $S$ wants $D$ to maximize the likelihood as
\begin{equation}
\label{eqn:seg_fake}  
    \mathcal{L}_{S_{{pred_{(x_i,\hat{y}_i)}}}} = - \mathbb{E}_{x_i,\hat{y}_i \sim S}\log[1 - D_i(x_i,\hat{y}_i)].
\end{equation} 
Since the main objective of the segmentor is to generate the segmentation map, $L_{S_{pred}}$ is usually weighed using a small number $\alpha$. 
In addition, a feature loss is calculated by collecting intermediate convolutions from the discriminators. The goal of feature matching is to push $S$ to generate segmentations that match reference data statistics. It is natural that $D$ can find the most discriminative features. The feature loss across all the discriminators is calculated and summed as
\begin{equation}
    \mathcal{L}_{S_{feature}} = \sum_i^d||f_{x,y\sim \mathcal{D}}(x_i,y_i) - f_{x,\hat{y}\sim S}(x_i,\hat{y}_i)||_2^2. 
\end{equation}

\paragraph{Discriminator Loss} The discriminator has only unsupervised loss objectives. When the model receives the stacked input image and reference segmentation label $(x,y)$ from two different sources, the unsupervised loss contains the original adversarial loss for real data: 
\begin{equation}
\label{eqn:D_real}  
    \mathcal{L}_{D_{i_{real}}} = - \mathbb{E}_{x_i,y_i \sim p_{data}} \log [ 1 - D_i(x_i,y_i)].
\end{equation}
Similarly, the adversarial loss for predicted data is calculated from the stacked input image and predicted segmentation label $x,\hat{y}$ as follows:
\begin{equation}
\label{eqn:D_fake}
\mathcal{L}_{D_{i_{pred}}} = - \mathbb{E}_{(x_i,\hat{y}_i) \sim S} \log [D_i(x_i,\hat{y}_i)].
\end{equation}

\paragraph{Encoder Loss}
The encoder is trained on matching the shapes of the predicted masks with the reference masks. The encoder $E$ is fed with both the reference mask and the predicted mask, and their latent representations are acquired as the outputs. The encoder loss is simply the mean-square error between the two latent representations: 
\begin{equation}
    \mathcal{L}_E = \frac{1}{n}\sum_i^n||z - \hat{z}||, 
\end{equation}
where $z$ is the latent vector representation of the reference segmentation and $\hat{z}$ is the latent vector representation of the predicted segmentation.

%%%%%%%%%%%Figure: Lung Segmentation
\begin{figure}
    \centering
    \resizebox{\linewidth}{!}{
    \begin{tabular}{ccccc}
    & {\Huge GT } & {\Huge MCU} & {\Huge JSRT} & {\Huge CHN}
    \smallskip
    \\
    {\Huge U-Net}
    \hspace{4pt}
    &
    \includegraphics[width=\linewidth, trim={4.75cm 0.25cm 4.75cm 1cm}, clip]{nlm_bound_031_nlm.pdf}
    &
    \includegraphics[width=\linewidth, trim={9.25cm 0.25cm 0.25cm 1cm}, clip]{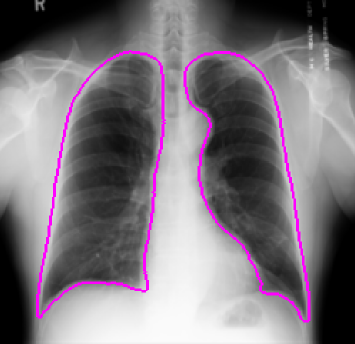}
        &
    \includegraphics[width=\linewidth, trim={9.25cm 0.25cm 0.25cm 1cm}, clip]{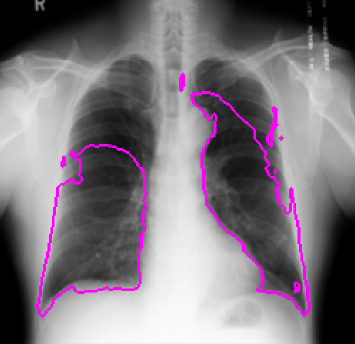}
    &
    \includegraphics[width=\linewidth, trim={9.25cm 0.25cm 0.25cm 1cm}, clip]{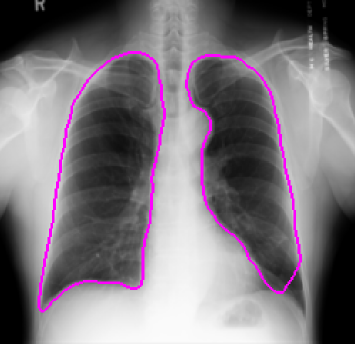}
    \\
    \smallskip
    \\
    &
    {\Huge PU-Net}
    \hspace{4pt}
    &
    \includegraphics[width=\linewidth, trim={9.25cm 0.25cm 0.25cm 1cm}, clip]{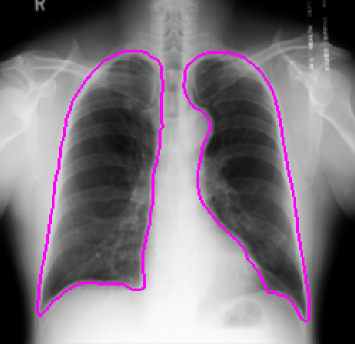}
        &
    \includegraphics[width=\linewidth, trim={9.25cm 0.25cm 0.25cm 1cm}, clip]{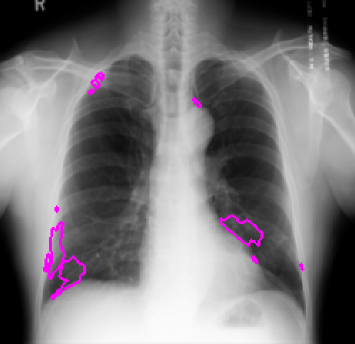}
    &
    \includegraphics[width=\linewidth, trim={9.25cm 0.25cm 0.25cm 1cm}, clip]{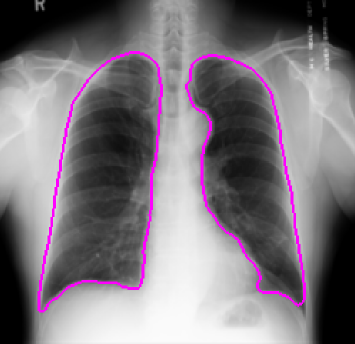}
    \\
    \smallskip
    \\
    &
    {\Huge ProgU-NetSS}
    \hspace{4pt}
    &
    \includegraphics[width=\linewidth, trim={9.25cm 0.25cm 0.25cm 1cm}, clip]{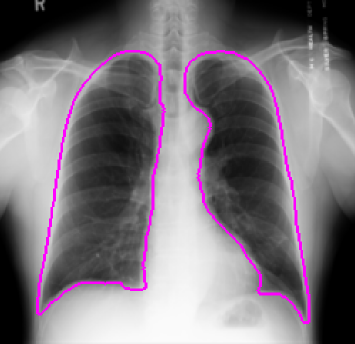}
        &
    \includegraphics[width=\linewidth, trim={9.25cm 0.25cm 0.25cm 1cm}, clip]{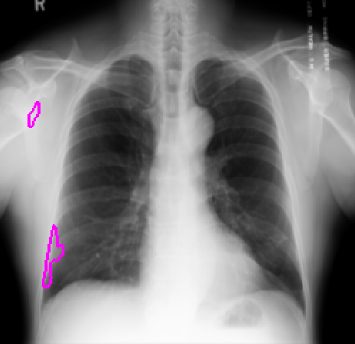}
    &
    \includegraphics[width=\linewidth, trim={9.25cm 0.25cm 0.25cm 1cm}, clip]{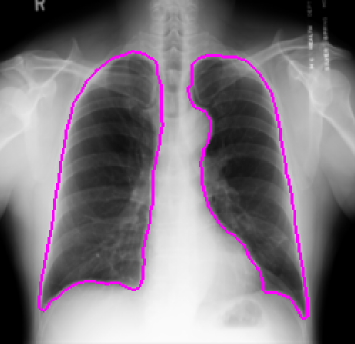}
    \\
    \smallskip
    \\
    &
    {\Huge AU-Net}
    \hspace{4pt}
    &
    \includegraphics[width=\linewidth, trim={9.25cm 0.25cm 0.25cm 1cm}, clip]{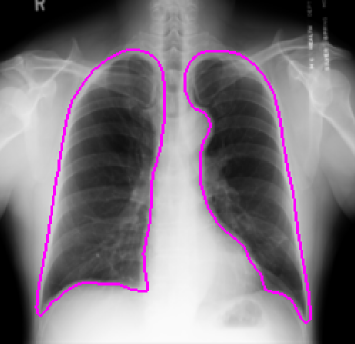}
        &
    \includegraphics[width=\linewidth, trim={9.25cm 0.25cm 0.25cm 1cm}, clip]{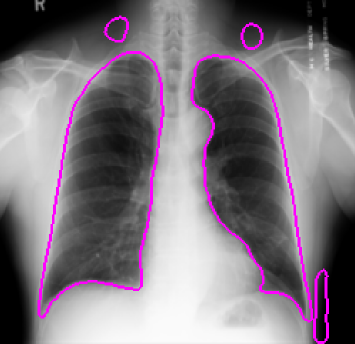}
    &
    \includegraphics[width=\linewidth, trim={9.25cm 0.25cm 0.25cm 1cm}, clip]{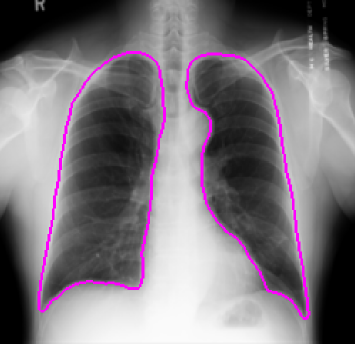}
    \\
    \smallskip
    \\
    &
    {\Huge APPU-Net}
    \hspace{4pt}
    &
    \includegraphics[width=\linewidth, trim={9.25cm 0.25cm 0.25cm 1cm}, clip]{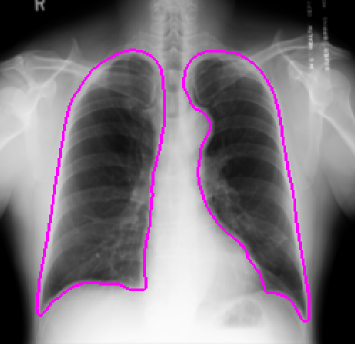}
        &
    \includegraphics[width=\linewidth, trim={9.25cm 0.25cm 0.25cm 1cm}, clip]{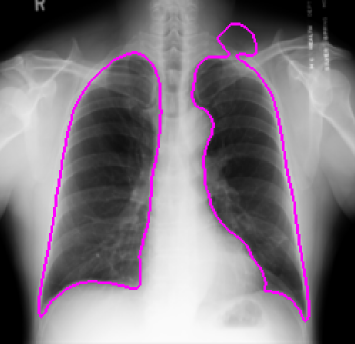}
    &
    \includegraphics[width=\linewidth, trim={9.25cm 0.25cm 0.25cm 1cm}, clip]{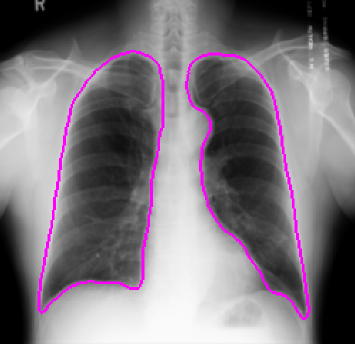}
    \\
    \smallskip
    &
    {\Huge PASS}
    \hspace{4pt}
    &
    \includegraphics[width=\linewidth, trim={9.25cm 0.25cm 0.25cm 1cm}, clip]{nlm_bound_031_nlm.pdf}
        &
    \includegraphics[width=\linewidth, trim={9.25cm 0.25cm 0.25cm 1cm}, clip]{nlm_bound_031_jsrt.pdf}
    &
    \includegraphics[width=\linewidth, trim={9.25cm 0.25cm 0.25cm 1cm}, clip]{nlm_bound_031_chn.pdf}
    \\
    \end{tabular}
   }
    \caption{Visualization of lung boundaries for the predicted segmentations of a chest X-Ray image from the MCU dataset when trained on the MCU, JSRT, and CHN datasets.}
    \label{fig:lung-nlm}
\end{figure}

%%%%%%%%%%%Figure: Vessel Segmentation
\begin{figure}
    \centering
    \resizebox{\linewidth}{!}{
    \begin{tabular}{ccccc}
   {\Huge Train} & {\Huge Input} & {\Huge GT } & {\Huge CHASE} & {\Huge ARIA}
    \smallskip
    \\
    {\Huge U-Net}
    \hspace{4pt}
    &
    \includegraphics[width=\linewidth, trim={0cm 0.25cm 9.5cm 0.75cm}, clip]{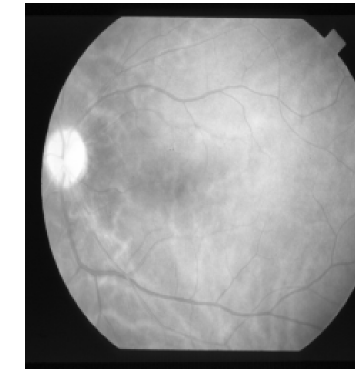}
    &
    \includegraphics[width=\linewidth, trim={4.75cm 0.25cm 4.75cm 0.75cm}, clip]{stare_002_chase_unet.pdf}
    &
    \includegraphics[width=\linewidth, trim={9.5cm 0.25cm 0cm 0.75cm}, clip]{stare_002_chase_unet.pdf}
    &
    \includegraphics[width=\linewidth, trim={9.5cm 0.25cm 0cm 0.75cm}, clip]{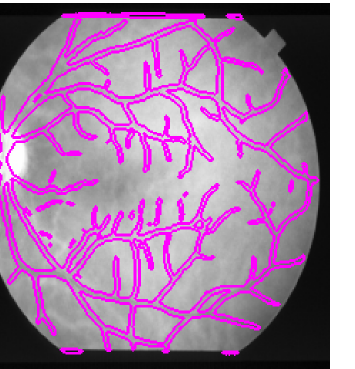}
    \\
    \smallskip
    \\
    &
    &
    {\Huge PU-Net}
    \hspace{4pt}
    &
    \includegraphics[width=\linewidth, trim={9.5cm 0.25cm 0cm 0.75cm}, clip]{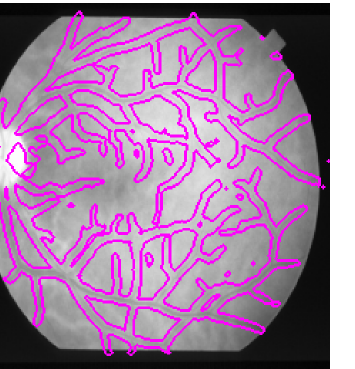}
    &
    \includegraphics[width=\linewidth, trim={9.5cm 0.25cm 0cm 0.75cm}, clip]{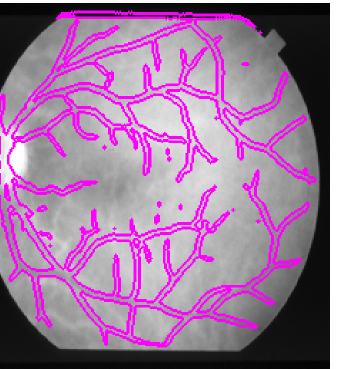}
    \\
    \smallskip
    \\
    &
    &
    {\Huge ProgU-NetSS}
    \hspace{4pt}
    &
    \includegraphics[width=\linewidth, trim={9.5cm 0.25cm 0cm 0.75cm}, clip]{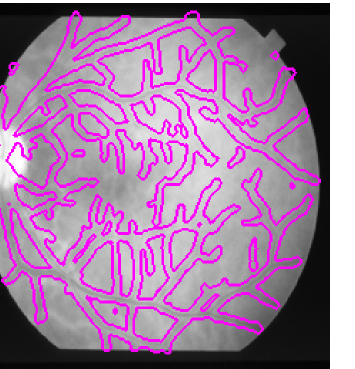}
    &
    \includegraphics[width=\linewidth, trim={9.5cm 0.25cm 0cm 0.75cm}, clip]{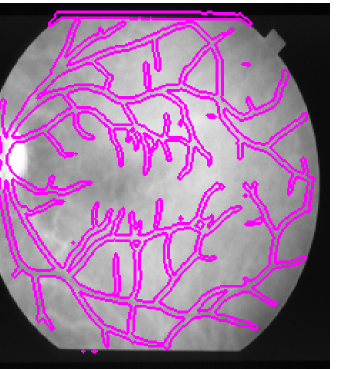}
    \\
    \smallskip
    \\
    &
    &
    {\Huge AU-Net}
    \hspace{4pt}
   &
    \includegraphics[width=\linewidth, trim={9.5cm 0.25cm 0cm 0.75cm}, clip]{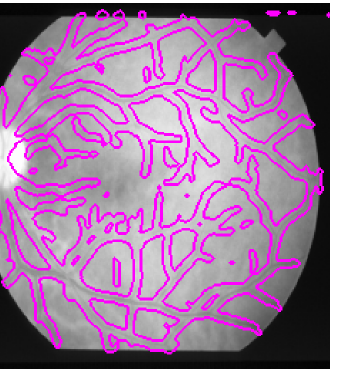}
    &
    \includegraphics[width=\linewidth, trim={9.5cm 0.25cm 0cm 0.75cm}, clip]{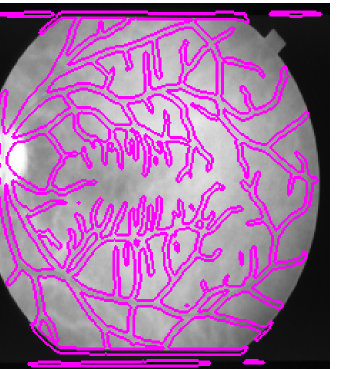}
    \\
    \smallskip
    \\
    &
    &
    {\Huge APPU-Net}
    \hspace{4pt}
    &
    \includegraphics[width=\linewidth, trim={9.5cm 0.25cm 0cm 0.75cm}, clip]{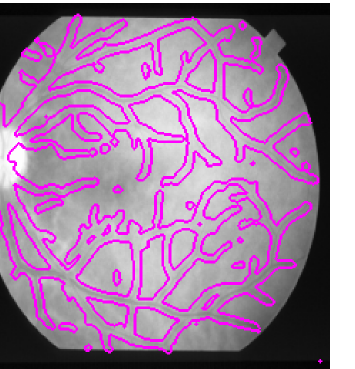}
    &
    \includegraphics[width=\linewidth, trim={9.5cm 0.25cm 0cm 0.75cm}, clip]{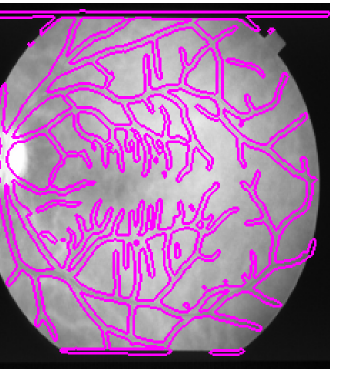}
    \\
    \smallskip
    &
    &
    {\Huge PASS}
    \hspace{4pt}
    &
    \includegraphics[width=\linewidth, trim={9.5cm 0.25cm 0cm 0.75cm}, clip]{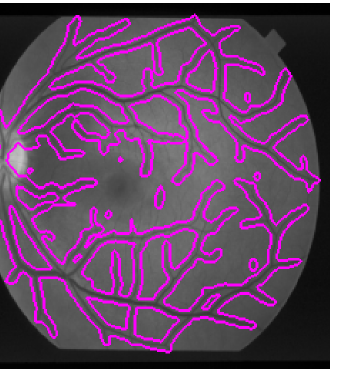}
    &
    \includegraphics[width=\linewidth, trim={9.5cm 0.25cm 0cm 0.75cm}, clip]{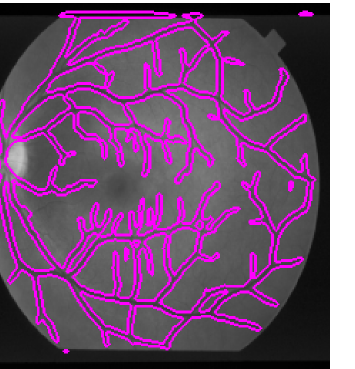}
    \\
    \end{tabular}
   }
    \caption{Visualization of vascular segmentation from fundoscopic image from the STARE dataset when the models are trained on the CHASE and ARIA datasets.}
    \label{fig:vessel-nlm}
\end{figure}

%%%%%Table-1: Dataset
\begin{table}
\setlength{\tabcolsep}{4pt}
\centering
\caption{Partitioning of the fundoscopic and chest X-ray datasets used in our experiments.}
\label{tab:dataset}
\resizebox{\linewidth}{!}{
\begin{tabular}{l c c c c c c ? l c c c c c c}
            \toprule
           \multirow{2}{*}{Dataset}
            &
           \multicolumn{3}{c}{\#Fundus images}
           &
           \multicolumn{3}{c?}{Data splits}
           &
           \multirow{2}{*}{Dataset}
           & 
           \multicolumn{3}{c}{\#X-ray images}
           &
           \multicolumn{3}{c}{Data splits}
           \\
           \cmidrule{2-7}\cmidrule{9-14}
            &
            Total & Healthy&Diseased 
            &
            Train & Val & Test
            &
            &
            Total & Healthy&Diseased 
            &
            Train & Val & Test
            \\
            \midrule
            \midrule
            DRIVE & 40 & 33 & 7 & 18 & 2 & 20
            &
            MCU & 138 & 80 & 58 & 93 & 10 & 35
            \\
            \midrule
            STARE & 20 & 10 & 10 & 10 & 2 & 8
            &
            JSRT & 247 & 93 & 154 & 111 & 13 & 123
            \\
            \midrule
            CHASE & 28 & 20 & 8 & 17 & 5 & 6
            &
            CHN & 566 & 279 & 287 & 377 & 43 & 142
            \\
            \midrule
            ARIA & 143 & 61 & 82 & 121 & 5 & 17
            \\
            \cmidrule{1-7}
            HRF & 45 & 15 & 30 & 26 & 5 & 14
            \\
            \cmidrule{1-7}
        \end{tabular}
}
\end{table}

\section{Empirical Evaluation}

\subsection{Dataset, Implementation, and Other Details}

\paragraph{Dataset} We validate our PASS model in two different experimental settings: vessel segmentation in diabetic retinopathy images and segmentation of lungs in chest X-ray images. For former task, we used five publicly available fundus image datasets---ARIA~\cite{bankhead2012fast}, CHASE~\cite{fraz2012ensemble}, DRIVE~\cite{staal2004ridge}, STARE~\cite{hoover2000locating}, and HRF~\cite{budai2013robust}--- with varying number of healthy and diseased images, and for the latter task, we used three public datasets---MCU, JSRT \cite{shiraishi2000development}, and CHN \cite{jaeger2014two}. Table~\ref{tab:dataset} details the partitioning of the datasets in each application. 

%%%%%Table-2: Vessel segmentation
\begin{table*}
\setlength{\tabcolsep}{4pt}
\centering
\caption{Comparison between PASS and other performance baselines for retinal vessel segmentation (Dice score). Grey columns indicate in-domain and white columns indicate cross-domain evaluations. 
% PASS outperforms the baselines and state-of-the-art models on-average, with a larger gain in the presence of domain shift and small sample size.
}
\label{tab:vessel-seg}
% \resizebox{\linewidth}{!}{
\begin{tabular}{l c a c c c c c ? c c a c c c c c c}
            \toprule
            \rowcolor{LightCyan}
            \multirow{2}{*}{Model} & Train on $\xrightarrow{}$ & \multicolumn{6}{c?}{CHASE}
            &
           \multicolumn{6}{c}{ARIA}
           \\
            \cmidrule{2-14}
           & Test on $\xrightarrow{}$ & CHASE & DRIVE & ARIA & STARE & HRF & Avg. 
           & CHASE & DRIVE & ARIA & STARE & HRF & Avg. 
            \\
            \midrule
            U-Net \cite{ronneberger2015u} & & 
            80.40 & 63.20 & 64.50 & 66.76 & 63.82 & 67.74 & 
            76.70 & 77.30 & 72.00 & 71.28 & 72.30 & 73.90
            \\
            \midrule
            U-Net+CRF \cite{tajbakhsh2019errornet}& & 
            81.20 & 65.40 & 62.60 & 56.40 & 63.60 & 65.80 & 
            78.40 & 69.50 & 73.00 & 64.60 & 73.50 & 71.80
            \\
            \midrule
            PU-Net \cite{fu2018joint} & & 
            81.58 & 64.04 & 63.03 & 66.20 & 62.66 & 67.50 & 
            76.7 & 77.3 & 72.0 & 71.28 & 72.3 & 73.9
            \\
            \midrule
            AttnU-Net \cite{oktay2018attention} & & 
            81.37 & 65.23 & 62.91 & 64.28 & 65.72 & 67.90 & 
            76.7 & 77.3 & 72.0 & 71.28 & 72.3 & 73.9 
            \\
            \midrule
            ProgU-Net \cite{imran2019end}& & 
            82.91 & 61.02 & 63.28 & 66.58 & 63.43 & 67.44 & 
            47.21 & 64.54 & 70.56 & 66.57 & 60.17 & 61.81
            \\
            \midrule
            ProgU-NetSS \cite{imran2019semi}& & 
            80.16 & 62.13 & 62.41 & 65.44 & 63.78 & 66.78 & 
            57.96 & 65.93 & 74.47 & 69.08 & 60.08 & 65.50
            \\
            \midrule
            V-GAN \cite{son2017retinal} & & 
            79.70 & 71.50 & 64.20 & 61.00 & 66.40 & 68.50 & 
            68.70 & 75.80 & 69.90 & 66.20 & 69.30 & 70.00
            \\
            \midrule
            AU-Net \cite{izadi2018generative} & & 
            82.20 & 63.20 & 61.84 & 67.17 & 63.37 & 67.56 & 
            68.06 & 70.21 & 78.12 & 74.95 & 69.69 & 72.21
            \\
            \midrule
            APPU-Net \cite{imran2019semi} & & 
            82.58 & 62.50 & 61.22 & 66.17 & 62.60 & 67.01 & 
            66.20 & 69.68 & 78.48 & 76.34 & 69.31 & 72.00
            \\
            \midrule
            UDA \cite{dong2018unsupervised} & & 
            72.30 & 69.30 & 68.20 & 64.70 & 67.40 & 68.40 & 
            71.50 & 72.90 & 73.20 & 71.30 & 70.70 & 71.90
            \\
            \midrule
            ErrorNet \cite{tajbakhsh2019errornet}& & 
            81.50 & 73.20 & 66.50 & 65.20 & 68.60 & 71.00 & 
            76.70 & 78.90 & 72.00 & 74.00 & 72.60 & 74.80
            \\
            \midrule
            PASS without $g(x)$ & & 
            89.06 & 80.76 & 80.72 & 82.72 & 75.26 & 81.70 &
            85.06 & 88.14 & 91.92 & 90.78 & 82.62 & 87.70
            \\
            \midrule
            PASS & & 
            {\bf91.96} & {\bf84.96} & {\bf84.18} & {\bf86.84} & {\bf78.57} & {\bf85.30} &
            {\bf86.32} & {\bf90.55} & {\bf92.08} & {\bf91.50} & {\bf83.15} & {\bf88.72}
            \\
            \bottomrule
        \end{tabular}
% }
\end{table*}

%%%%%Table-2: Lung segmentation
\begin{table*}
\setlength{\tabcolsep}{4pt}
\centering
\caption{Comparison between PASS and other performance baselines for lung segmentation from chest X-ray images (Dice score). Grey columns indicate in-domain evaluation and white columns indicate cross-domain evaluations.}
\label{tab:lung-seg}
% \resizebox{\linewidth}{!}{
\begin{tabular}{l c a c c c ? c a c c ? c c a c}
            \toprule
             \rowcolor{LightCyan}
             \multirow{2}{*}{Model}
             &
            Train on $\xrightarrow{}$ & \multicolumn{4}{c?}{MCU} & \multicolumn{4}{c?}{JSRT} & \multicolumn{4}{c}{CHN} 
           \\
           \cmidrule{2-14}
           & Test on $\xrightarrow{}$ & MCU & JSRT & CHN & Avg. 
           & MCU & JSRT & CHN & Avg. & MCU & JSRT & CHN & Avg. 
            \\
            \midrule
            U-Net \cite{ronneberger2015u}& & 
            97.67 & 39.39 & 94.48 & 77.18 & 
            92.00 & 95.02 & 90.54 & 92.58 &
            93.72 & 43.46 & 95.84 & 77.67
            \\
            \midrule
            PU-Net \cite{fu2018joint}& & 
            97.89 & 21.24 & {\bf97.84} & 72.33 & 
            84.97 & 94.94 & 73.68 & 84.53 & 
            93.57 & 73.88 & 95.90 & 87.78
            \\
            \midrule
            AttnU-Net \cite{oktay2018attention}& & 
            97.86 & 30.31 & 94.07 & 74.08 & 
            6.70 & 94.95 & 65.00 & 55.55 & 
            81.25 & 74.24 & 95.56 & 83.68
            \\
            \midrule
            ProgU-Net \cite{imran2019end} & & 
            97.83 & 10.98 & 91.32 & 66.71 & %MCU
            34.89 & 95.20 & 86.28 & 72.12 &  %jsrt
            84.79 & 60.03 & 95.35 & 80.06 %CHN
            \\
            \midrule
            ProgU-NetSS  \cite{imran2019semi} & & 
            97.90 & 33.98 &  95.32 & 75.33 & 
            13.16 & 95.09 & 65.00 & 57.75 & 
            94.24 & 67.29 & 95.63 & 85.72 
            \\
            \midrule
            AU-Net \cite{izadi2018generative}& & 
            97.86 & 94.68 & 95.08 & 95.87 & 
            89.12 & 97.85 & 92.46 & 93.14 & 
            95.58 & 95.88 & 96.22 & 95.89
            \\
            \midrule
            APPU-Net \cite{imran2019semi} & & 
            97.81 &  95.07 & 94.77 & 95.88 & 
            90.46 & 97.80 & 91.76 & 93.34 &
            95.72 & 96.25 & 96.11 & 96.03
            \\
            \midrule
            CyUDA \cite{chen2018semantic} & &
            95.61 & 92.84 & -- & 94.23 &
            -- & -- & -- & -- &
            -- & -- & -- & --
            \\
            
            \midrule
            SeUDA \cite{chen2018semantic} & &
            95.61 & 94.51 & -- & --
            -- & -- & -- & -- &
            -- & -- & -- & --
            \\
            \midrule
            CoDAGAN \cite{oliveira2019truly} & & 
            -- & -- & -- & -- & 
            84.58 & 96.45 & 88.99 & 90.01 & 
            -- & -- & -- & -- 
            \\
            \midrule
            PASS without $g(x)$ & & 
            97.74 & 96.43 & 96.76 & 96.98 & 
            95.11 & 98.26 & 95.92 & 96.43 &
            96.62 & 96.11 & 97.61 & 96.68
            \\
            \midrule
            PASS & & 
            {\bf98.22} & {\bf96.56} & 97.24 & {\bf97.34} & 
            {\bf95.70} & {\bf98.27} & {\bf 96.06} & {\bf96.68} & 
            {\bf97.27} & {\bf97.15} & {\bf97.65} & {\bf97.36}
            \\
            \bottomrule
        \end{tabular}
% }
\end{table*}

\paragraph{Inputs} All the images are resized and normalized to $256\times256\times3$ for the retinal images and $256\times256\times1$ for the chest X-rays before feeding them to the network.

\paragraph{Hyperparameters} We use the Adam optimizer with adaptive learning rate starting with initial rates of $0.01$ for $S$, and $0.001$ for $D$ and $E$. The learning rates are decreased 90\% after every 5 epochs with exponential decay. We apply the dropout with a rate of 0.25. We use $\lambda = 0.3$ for weighting $\mathcal{L_A}$ and $\alpha = 0.01$ in (\ref{eqn:seg_fake}). 

\paragraph{Machine Configuration} We implemented PASS in Tensorflow on a Nvidia Titan V GPU and a 64-bit Intel(R) Core(TM) i7-9700K CPU. 

\paragraph{Evaluation} Along with qualitative visualization of segmentation masks and edges overlaid on the original input images, we use the Dice score (DS), structural similarity index (SSIM), and average Hausdorff distance (HD).

\paragraph{Baselines} We employ a number of baseline and state-of-the-art models for medical image segmentation and domain adaptation.\footnote{U-Net, U-Net with CRF, Pyramid U-Net (PU-Net), Progressive U-Net (ProgU-Net), Attention U-Net (AttnU-Net), Progressive U-Net with Side-Supervision (ProgU-NetSS), Adversarial U-Net (AU-Net), V-GAN, Adversarial Pyramid Progressive U-Net (APPU-Net), Unsupervised Domain Adaptation (UDA), ErrorNet, and Conditional Domain Adaptation with GAN (CoDAGAN).}

\subsection{Results}

As reported in Tables~\ref{tab:vessel-seg} and \ref{tab:lung-seg}, PASS outperforms all the baselines and state-of-the-art models for both retinal vessel and lung segmentation tasks. 

%%%%%%%%%%%%%%%%%%%%%%%%%%%%%%%%%%%%%%%%

%%%%%Table: Lung segmentation
\begin{table*}
\setlength{\tabcolsep}{4pt}
\centering
\caption{Comparison between PASS and other performance baselines for pulmonary segmentation (HD score). Grey columns indicate in-domain and white columns indicate cross-domain evaluations.}
\label{tab:hd-lung}
% \resizebox{\linewidth}{!}{
\begin{tabular}{l c a c c c ? c a c c ? c c a c}
            \toprule
             \rowcolor{LightCyan}
             \multirow{2}{*}{Model}
             &
            Train on $\xrightarrow{}$ & \multicolumn{4}{c?}{MCU} & \multicolumn{4}{c?}{JSRT} & \multicolumn{4}{c}{CHN} 
           \\
           \cmidrule{2-14}
           & Test on $\xrightarrow{}$ & MCU & JSRT & CHN & Avg. 
           & MCU & JSRT & CHN & Avg. & MCU & JSRT & CHN & Avg. 
            \\
            \midrule
            U-Net & & 
            3.689 & 9.948 & 4.849 & 6.162 &
            7.923 & 4.128 & 5.594 & 5.882 &
            4.874 & 9.161 & 4.378 & 6.138
            \\
            \midrule
            PU-Net & & 
           3.723 & 11.049 & 5.346 & 6.706 &
           10.926 & 4.159 & 6.618 & 7.234 &
           5.347 & 7.438 & 4.362 & 5.716
            \\
            \midrule
            AttnU-Net & & 
            3.841 & 10.189 & 5.057 & 6.362 &
            11.628 & 4.181 & 7.374 & 7.728 &
            6.989 & 7.257 & 4.453 & 6.233
            \\
            \midrule
            ProgU-Net & & 
            3.729 & 11.706 & 5.564 & 6.999 &
            10.491 & 4.052 & 5.700 & 6.748 &
            6.788 & 8.702 & 4.494 & 6.661
            \\
            \midrule
            ProgU-NetSS  & & 
           3.874 & 9.907 & 4.673 & 6.151 &
           11.602 & 4.117 & 7.509 & 7.743 &
           4.885 & 7.906 & 4.362 & 5.718
            \\
            \midrule
            AU-Net & & 
            3.780 & 4.625 & 4.651 & 4.352 &
            4.879 & 3.705 & 5.126 & 4.570 &
            4.327 & 4.499 & 4.301 & 4.376
            \\
            \midrule
            APPU-Net & & 
            3.736 & 4.537 & 4.725 & 4.333 &
            4.732 & 3.706 & 5.055 & 4.498 &
            4.485 & 4.278 & 4.401 & 4.388
            \\
            \midrule
            PASS without $g(x)$ & & 
            3.332 & 4.552 & 4.857 & 4.247 &
            5.109 & 3.867 & 5.138 & 4.705 &
            4.580 & 4.441 & 4.545 & 4.522
            \\
            \midrule
            PASS & & 
            3.262 & 3.996 & 4.675 & 3.978 &
            4.680 & 3.702 & 5.021 & 4.468 &
            4.269 & 4.397 & 4.112 &  4.259
            \\
            \bottomrule
        \end{tabular}
% }
\end{table*}

%%%%%Table-2: Lung segmentation
\begin{table*}
\setlength{\tabcolsep}{4pt}
\centering
\caption{Comparison between PASS and other performance baselines for pulmonary segmentation (SSIM score). Grey columns indicate in-domain and white columns indicate cross-domain evaluations.}
\label{tab:ssim-lung}
% \resizebox{\linewidth}{!}{
\begin{tabular}{l c a c c c ? c a c c ? c c a c}
            \toprule
             \rowcolor{LightCyan}
             \multirow{2}{*}{Model}
             &
            Train on $\xrightarrow{}$ & \multicolumn{4}{c?}{MCU} & \multicolumn{4}{c?}{JSRT} & \multicolumn{4}{c}{CHN} 
           \\
           \cmidrule{2-14}
           & Test on $\xrightarrow{}$ & MCU & JSRT & CHN & Avg. 
           & MCU & JSRT & CHN & Avg. & MCU & JSRT & CHN & Avg. 
            \\
            \midrule
            U-Net & & 
            0.968 & 0.754 & 0.935 & 0.886 & 
            0.803 & 0.949 & 0.908 & 0.887 &
            0.929 & 0.738 & 0.949 & 0.872
            \\
            \midrule
            PU-Net & & 
            0.967 & 0.672 & 0.921 & 0.853 & 
            0.698 & 0.948 & 0.839 & 0.828 &
            0.931 & 0.836 & 0.949 & 0.905
            \\
            \midrule
            AttnU-Net & & 
            0.967 & 0.699 & 0.926 & 0.864 &
            0.679 & 0.948 & 0.820 & 0.816 &
            0.869 & 0.837 & 0.946 & 0.884
            \\
            \midrule
            ProgU-Net & & 
            0.968 & 0.656 & 0.919 & 0.848 &
            0.732 & 0.951 & 0.889 & 0.857 &
            0.886 & 0.837 & 0.946 & 0.889
            \\
            \midrule
            ProgU-NetSS  & & 
            0.968 & 0.716 & 0.941 & 0.875 &
            0.689 & 0.949 & 0.828 & 0.822 &
            0.934 & 0.822 & 0.948 & 0.901 
            \\
            \midrule
            AU-Net & & 
            0.967 & 0.9191 & 0.936 & 0.941 &
            0.892 & 0.961 & 0.916 & 0.923 &
            0.942 & 0.937 & 0.951 & 0.943
            \\
            \midrule
            APPU-Net & & 
            0.967 & 0.922 & 0.934 & 0.941 &
            0.902 & 0.960 & 0.913 & 0.925 &
            0.944 & 0.940 & 0.950 & 0.945
            \\
            \midrule
            PASS without $g(x)$ & & 
            0.962 & 0.921 & 0.932 & 0.938 &
            0.913 & 0.953 & 0.921 & 0.929 &
            0.939 & 0.931 & 0.945 & 0.938 
            \\
            \midrule
            PASS & & 
            0.968 & 0.923 & 0.942 & 0.944 &
            0.918 & 0.955 & 0.920 & 0.931 &
            0.942 & 0.939 & 0.956 & 0.946
            \\
            \bottomrule
        \end{tabular}
% }
\end{table*}

In vessel segmentation, PASS achieved an overall average Dice score of 85.30 and cross-domain score of 83.74 (domain gap of 8.32), when trained on the small CHASE dataset. When trained on the relatively larger ARIA dataset, the overall average Dice score of 88.72 and cross-domain score of 87.88 are achieved with domain gap of 4.22. This demonstrates the effectiveness of PASS across the domains as it is capable of performing the semantic segmentation with the ability to learn diversity in shapes and distributions. 
Similarly, in the segmentation of lungs, PASS outperformed all the baselines and state-of-the-art models (Table~\ref{tab:lung-seg}). 

%%%%%%%%%%%%%%%%%%%%%%%%%%%%%%%%%%%%%%%%%%%%%%%%%%%%

%%%%%Table-1: HD: Vessel segmentation
\begin{table*}
\setlength{\tabcolsep}{4pt}
\centering
\caption{Comparison between PASS and other performance baselines for retinal vessel segmentation (average HD score). Grey columns indicate in-domain and white columns indicate cross-domain evaluations.}
\label{tab:hd-retina}
% \resizebox{\linewidth}{!}{
\begin{tabular}{c c a c c c c c ? c c a c c c c c c}
            \toprule
            \rowcolor{LightCyan}
            \multirow{2}{*}{Model} & Train on $\xrightarrow{}$ & \multicolumn{6}{c?}{CHASE}
            &
           \multicolumn{6}{c}{ARIA}
           \\
            \cmidrule{2-14}
           & Test on $\xrightarrow{}$ & CHASE & DRIVE & ARIA & STARE & HRF & Avg. 
           & CHASE & DRIVE & ARIA & STARE & HRF & Avg. 
            \\
            \midrule
            U-Net & & 
            7.929 & 8.306 & 7.490 & 7.807 & 8.916 & 8.089 &
            9.177 & 8.095 & 6.404 & 7.209 & 9.017 & 7.980 
            \\
            \midrule
            PU-Net & & 
            7.287 & 8.260 & 7.526 & 7.551 & 8.815 & 7.887 &
            9.381 & 8.247 & 6.905 & 7.420 & 9.403 & 8.271
            \\
            \midrule
            AttnU-Net & & 
            7.436 & 8.515 & 7.672 & 7.899 & 8.982 & 8.100 &
            9.765 & 8.297 & 7.186 & 7.691 & 9.543 & 8.496
            \\
            \midrule
            ProgU-Net & & 
            7.233 & 8.408 & 7.613 & 7.850 & 9.000 & 8.021 &
            9.400 & 8.144 & 6.627 & 7.316 & 9.133 & 8.124
            \\
            \midrule
            ProgU-NetSS & & 
            7.498 & 8.478 & 7.560 & 8.055 & 9.012 & 8.121 &
            8.857 & 8.072 & 6.478 & 7.089 & 9.143 & 7.928
            \\
            \midrule
            AU-Net & & 
            7.406 & 8.482 & 7.665 & 7.792 & 8.946 & 8.058 &
            8.508 & 7.770 & 6.448 & 6.606 & 8.510 & 7.568
            \\
            \midrule
            APPU-Net & & 
            7.196 & 8.492 & 7.732 & 7.824 & 8.979 & 8.045 &
            8.486 & 8.009 & 6.344 & 6.519 & 8.655 & 7.602
            \\
            \midrule
            PASS without $g(x)$ & & 
            7.154 & 7.809 & 7.397 & 7.346 & 8.261 & 7.593 &  
            8.349 & 8.132 & 6.301 & 6.120 & 8.074 & 7.396
            \\
            \midrule
            PASS & & 
            7.129 & 7.341 & 7.187 & 7.745 & 8.028 & 7.486 &
            8.285 & 8.012 & 6.336 & 6.123 & 7.399 & 7.231

            \\
            \bottomrule
        \end{tabular}
% }
\end{table*}

%%%%%%%%%%%%%%%%%%%%%%%%%%%%%%%%%%%%%%%%%%%%%%%%

%%%%%Table-1: SSIM: Vessel segmentation
\begin{table*}
\setlength{\tabcolsep}{4pt}
\centering
\caption{Comparison between PASS and other performance baselines for retinal vessel segmentation (SSIM score). Grey columns indicate in-domain and white columns indicate cross-domain evaluations.}
\label{tab:ssim-retina}
% \resizebox{\linewidth}{!}{
\begin{tabular}{c c a c c c c c ? c c a c c c c c c}
            \toprule
            \rowcolor{LightCyan}
            \multirow{2}{*}{Model} & Train on $\xrightarrow{}$ & \multicolumn{6}{c?}{CHASE}
            &
           \multicolumn{6}{c}{ARIA}
           \\
            \cmidrule{2-14}
           & Test on $\xrightarrow{}$ & CHASE & DRIVE & ARIA & STARE & HRF & Avg. 
           & CHASE & DRIVE & ARIA & STARE & HRF & Avg. 
            \\
            \midrule
            U-Net & & 
            0.812 & 0.645 & 0.665 & 0.684 & 0.507 & 0.663 & 
            0.682 & 0.734 & 0.791 & 0.776 & 0.571 & 0.711
            \\
            \midrule
            PU-Net & & 
            0.815 & 0.678 & 0.666 & 0.699 & 0.501 & 0.672 &
            0.679 & 0.722 & 0.769 & 0.751 & 0.537 & 0.692
            
            \\
            \midrule
            AttnU-Net & & 
            0.808 & 0.645 & 0.648 & 0.687 & 0.498 & 0.657 &
            0.675 & 0.725 & 0.764 & 0.753 & 0.543 & 0.692
           \\
            \midrule
            ProgU-Net & & 
            0.819 & 0.611 & 0.653 & 0.679 & 0.490 & 0.650 & 
            0.677 & 0.739 & 0.791 & 0.774 & 0.595 & 0.715
         
            \\
            \midrule
            ProgU-NetSS & & 
            0.802 & 0.624 & 0.652 & 0.659 & 0.501 & 0.648 &
            0.698 & 0.741 & 0.797 & 0.783 & 0.586 & 0.721
           
            \\
            \midrule
            AU-Net & & 
            0.812 & 0.624 & 0.641 & 0.658 & 0.482 & 0.643 &
            0.706 & 0.738 & 0.789 & 0.779 & 0.632 & 0.729 
        
            \\
            \midrule
            APPU-Net & & 
            0.820 & 0.619 & 0.636 & 0.654 & 0.484 & 0.643 &
            0.695 & 0.738 & 0.796 & 0.789 & 0.632 & 0.730
    
            \\
            \midrule
            PASS without $g(x)$ & & 
            0.827 & 0.682 & 0.661 & 0.699 & 0.505 & 0.675 &
            0.696 & 0.742 & 0.798 & 0.791 & 0.610 & 0.725
            \\
            \midrule
            PASS & & 
            0.830 & 0.685 & 0.660 & 0.701 & 0.510 & 0.677 &
            0.707 & 0.772 & 0.797 & 0.802 & 0.611 & 0.738
            \\
            \bottomrule
        \end{tabular}
% }
\end{table*}

Along with Fig.~\ref{fig:teaser}, Fig.~\ref{fig:lung-nlm} (edges) shows the consistency of PASS in segmenting lungs from the chest X-ray  
and Fig.~\ref{fig:vessel-nlm} shows the consistency of PASS in segmenting retinal vessels from both in-domain and cross-domains irrespective of the varying imaging configurations and abnormalities.
While some of the baseline models completely failed in evaluations on the JSRT dataset when trained on MCU and vice-versa, PASS consistently performs better in either scenario. With PASS, we have a domain gap of only 1.32 when trained on MCU, 2.39 when trained on JSRT, and 0.44 when trained on CHN. The poorer performance of the PASS model without $g(x)$ justifies the inclusion of this transformation function and the logit-wise distribution matching.

\iffalse
\begin{figure}
\centering
 \resizebox{\linewidth}{!}{%
  \begin{tabular}{cccccc}
    \includegraphics[width=\linewidth, trim={0.5cm 0.5cm 0.5cm 0cm},clip]{bound_003_chase.pdf}
    \end{tabular} }
    \medskip
    \caption{Boundary visualization of the predicted vertebrae masks in a spine X-Ray shows consistent improvement by our proposed model over all other models.}
    \label{fig:edge}
\end{figure}
\fi

The HD and SSIM scores are reported for lung segmentation in
Table~\ref{tab:hd-lung} and
Table~\ref{tab:ssim-lung}, and for retinal vessel segmentation in
Table~\ref{tab:hd-retina} and Table~\ref{tab:ssim-retina}. The consistently lower HD scores and higher SSIM scores compared to the baselines provide further evidence of the superior performance of our PASS model.

\section{Conclusions}

We introduced PASS, a novel semantic segmentation model for intelligently
mitigating the domain shift problem caused by small datasets. We evaluated PASS using 8 public datasets for the tasks of retinal vessel segmentation from diabetic retinopathy images and lung segmentation from chest X-ray images. Our experimental results demonstrated the effectiveness of PASS in both in-domain and cross-domain evaluations, even with smaller sample size and larger domain shift. Our future work will focus on evaluating PASS on other medical image segmentation tasks as well as evaluating the effectiveness of PASS in iterative and continual learning scenarios.

\balance
\bibliographystyle{IEEEtran}
\bibliography{refs}

\clearpage

\section*{Supplementary Material}

% The details of the components of the PASS model architecture are presented in the tables below:

\begin{table}[!htb]
\setlength{\tabcolsep}{4pt}
    \centering
     \caption{architectural details of the shape encoder $(E)$}
    \label{tab:enc}
    \resizebox{\linewidth}{!}{
    \begin{tabular}{c c c}

        \toprule
        Name & Feature maps (input) & Feature maps (output) 
        \\
        \midrule
        Conv layer - 1a & $256\times256\times3$ & $256\times256\times16$ 
        
        \\
        Conv layer - 1b & $256\times256\times16$ & $256\times256\times16$ 
        
        \\
        Max pool - 1 & $256\times256\times16$ & $128\times128\times16$ 
        
        \\
        Conv layer - 2a & $128\times128\times16$ & $128\times128\times32$ 
        
        \\
        Conv layer - 2b & $128\times128\times32$ & $128\times128\times32$ 
        
        \\
        Max pool - 2 & $128\times128\times32$ & $64\times64\times32$ 
        
        \\
        Conv layer - 3a & $64\times64\times32$  & $64\times64\times64$ 
        
        \\
        Conv layer - 3b & $64\times64\times64$ & $64\times64\times64$  
        
        \\
        Max pool - 3 & $64\times64\times64$  & $32\times32\times64$ 
        
        \\
        Conv layer - 4a & $32\times32\times64$  & $32\times32\times128$  
        
        \\
        Conv layer - 4b & $32\times32\times128$  & $32\times32\times128$  
        
        \\
        Max pool - 4 & $32\times32\times128$ & $16\times16\times128$
        
        \\
        Conv layer - 5a & $16\times16\times128$  & $16\times16\times256$  
        
        \\
        Conv layer - 5b & $16\times16\times256$  & $16\times16\times256$  
        
        \\
        encoder flatten - 5 & $16\times16\times256$  & 65536
        
        \\
        encoder dense - $z$ & 65536  & 256
        \\
        \bottomrule
    \end{tabular}
    }
\end{table}

\begin{table}[!htb]
\setlength{\tabcolsep}{4pt}
    \centering
     \caption{architectural details of the Discriminator $(D)$}
    \label{tab:enc}
    \resizebox{\linewidth}{!}{
    \begin{tabular}{c c c}
        \toprule
        Name & Feature maps (input) & Feature maps (output) 
        \\
        \midrule
        Conv layer - 1a & $256\times256\times3$ & $256\times256\times16$ 
        
        \\
        Conv layer - 1b & $256\times256\times16$ & $256\times256\times16$ 
        
        \\
        Max pool - 1 & $256\times256\times16$ & $128\times128\times16$ 
        
        \\
        Conv layer - 2a & $128\times128\times16$ & $128\times128\times32$ 
        
        \\
        Conv layer - 2b & $128\times128\times32$ & $128\times128\times32$ 
        
        \\
        Max pool - 2 & $128\times128\times32$ & $64\times64\times32$ 
        
        \\
        Conv layer - 3a & $64\times64\times32$  & $64\times64\times64$ 
        
        \\
        Conv layer - 3b & $64\times64\times64$ & $64\times64\times64$  
        
        \\
        Max pool - 3 & $64\times64\times64$  & $32\times32\times64$ 
        
        \\
        Conv layer - 4a & $32\times32\times64$  & $32\times32\times128$  
        
        \\
        Conv layer - 4b & $32\times32\times128$  & $32\times32\times128$  
        
        \\
        Max pool - 4 & $32\times32\times128$ & $16\times16\times128$
        
        \\
        discriminator flatten - 4 & $16\times16\times128$  &  32768
        
        \\
        discriminator dense - $l$ & 32768  & 1
        \\
        \bottomrule
    \end{tabular}
    }
\end{table}

\begin{table}[!h]
\setlength{\tabcolsep}{4pt}
    \centering
     \caption{architectural details of the Discriminator $(D_2)$}
    \label{tab:enc}
    \resizebox{\linewidth}{!}{
    \begin{tabular}{c c c}
        \toprule
        Name & Feature maps (input) & Feature maps (output) 
        \\
        \midrule
        Conv layer - 1a & $128\times128\times3$ & $128\times128\times32$ 
        
        \\
        Conv layer - 1b & $128\times128\times32$ & $128\times128\times32$ 
        
        \\
        Max pool - 1 & $128\times128\times32$ & $64\times64\times32$ 
        
        \\
        Conv layer - 2a & $64\times64\times32$  & $64\times64\times64$ 
        
        \\
        Conv layer - 2b & $64\times64\times64$ & $64\times64\times64$  
        
        \\
        Max pool - 2 & $64\times64\times64$  & $32\times32\times64$ 
        
        \\
        Conv layer - 3a & $32\times32\times64$  & $32\times32\times128$  
        
        \\
        Conv layer - 3b & $32\times32\times128$  & $32\times32\times128$  
        
        \\
        Max pool - 3 & $32\times32\times128$ & $16\times16\times128$
        
        \\
        discriminator flatten - 3 & $16\times16\times128$  &  32768
        
        \\
        discriminator dense - $l$ & 32768  & 1
        \\
        \bottomrule
    \end{tabular}
    }
\end{table}

\begin{table}[!hb]
\setlength{\tabcolsep}{4pt}
    \centering
     \caption{architectural details of the Discriminator $(D_4)$}
    \label{tab:enc}
    \resizebox{\linewidth}{!}{
    \begin{tabular}{c c c}
        \toprule
        Name & Feature maps (input) & Feature maps (output)
        \\
        \midrule
        Conv layer - 1a & $64\times64\times3$  & $64\times64\times64$ 
        
        \\
        Conv layer - 1b & $64\times64\times64$ & $64\times64\times64$  
        
        \\
        Max pool - 1 & $64\times64\times64$  & $32\times32\times64$ 
        
        \\
        Conv layer - 2a & $32\times32\times64$  & $32\times32\times128$  
        
        \\
        Conv layer - 2b & $32\times32\times128$  & $32\times32\times128$  
        
        \\
        Max pool - 2 & $32\times32\times128$ & $16\times16\times128$
        
        \\
        discriminator flatten - 2 & $16\times16\times128$  &  32768
        
        \\
        discriminator dense - $l$ & 32768  & 1
        \\
        \bottomrule
    \end{tabular}
    }
\end{table}

\begin{table}[!hb]
\setlength{\tabcolsep}{4pt}
    \centering
     \caption{architectural details of the Discriminator $(D_8)$}
    \label{tab:enc}
    \resizebox{\linewidth}{!}{
    \begin{tabular}{c c c}
        \toprule
        Name & Feature maps (input) & Feature maps (output)
        \\
        \midrule
        Conv layer - 1a & $32\times32\times3$  & $32\times32\times128$  
        
        \\
        Conv layer - 1b & $32\times32\times128$  & $32\times32\times128$  
        
        \\
        Max pool - 1 & $32\times32\times128$ & $16\times16\times128$
        
        \\
        discriminator flatten - 1 & $16\times16\times128$  &  32768
        
        \\
        discriminator dense - $l$ & 32768  & 1
        \\
        \bottomrule
    \end{tabular}
    }
\end{table}

\end{document}